\documentclass[journal]{IEEEtran}

\usepackage{cite}
\usepackage{amsmath,amssymb,amsfonts}
\usepackage{algorithmic}
\usepackage{graphicx}
\usepackage{textcomp}
\usepackage{xcolor}
\usepackage{multicol}
\usepackage{booktabs}
\usepackage{multirow}
\usepackage{gensymb}

\usepackage{soul}
\usepackage{graphicx}
\usepackage{amsmath}
\usepackage{booktabs}
\usepackage{threeparttable}
\usepackage[hidelinks]{hyperref}

\begin{document}

\title{Design and Implementation of a Secure RISC-V Microprocessor}

\author{\IEEEauthorblockN{Kleber Stangherlin, Manoj Sachdev} \\
        \IEEEauthorblockA{ECE Department, University of Waterloo, Waterloo, ON N2L 3G1, Canada}\\
        \{khstangh,msachdev\}@uwaterloo.ca
}

\markboth{DOI: 10.1109/TVLSI.2022.3203307 \quad Copyright \copyright 2022 IEEE}{}

\maketitle

\begin{abstract}
Secret keys can be extracted from the power consumption or electromagnetic emanations of unprotected devices. Traditional counter-measures have limited scope of protection, and impose several restrictions on how sensitive data must be manipulated. We demonstrate a bit-serial RISC-V microprocessor implementation with no plain-text data. All values are protected using Boolean masking. Software can run with little to no counter-measures, reducing code size and performance overheads. Unlike previous literature, our methodology is fully automated and can be applied to designs of arbitrary size or complexity. We also provide details on other key components such as clock randomizer, memory protection, and random number generator. The microprocessor was implemented in 65~nm CMOS technology. Its implementation was evaluated using NIST tests as well as side channel attacks. Random numbers generated with our RNG pass on all NIST tests. Side-channel analysis on the baseline implementation extracted the AES key using only 375 traces, while our secure microprocessor was able to withstand attacks using 20~M traces.
\end{abstract}


\section{Introduction}

Increased cyber crime and decentralized currencies are escalating the importance of protecting digital assets. Malicious individuals use a number of techniques to extract secrets from integrated circuits (ICs). Traditional hardware security counter-measures only target a few most vulnerable elements, such as system bus, memory, and cryptographic accelerators. We demonstrate a bit-serial RISC-V microprocessor implementation with no plain-text data. All values are protected using Boolean masking. Our architecture sets no constraints on how sensitive data must be manipulated. Software implementations may run with little to no counter-measures, reducing code size and performance overheads.

Unlike previous literature, our methodology is fully integrated to the ASIC digital design flow. We carefully selected a set of counter-measures that require no changes to the input register transfer level (RTL) code, or to the application software. Our techniques can be applied to digital designs of any size or complexity. Original circuit functionality and latency are not affected.

Our design uses Boolean masking (BM) at the logic level. Boolean masking splits each value into two or more shares. Individual shares are (ideally) uncorrelated to the original value. Physical probing of all shares is required to uncover the original data. Accurate tampering of the system requires precise modifications in all shares of a value, which is considerably more difficult.

At the transistor level, we use differential domino logic (DDL) to implement our logic gates. DDL is a differential input/output dynamic logic style with precharge/evaluation phases. During precharge, both outputs are driven to the same value, but only one of them toggles in the evaluation phase. DDL is used to suppress glitches and reduce the data-dependent power consumption.

To allow meaningful comparison, we implemented three different version of a bit-serial RISC-V microprocessor using the same RTL code as input. A baseline implementation with no counter-measures, another using Boolean masking, and a third that combines Boolean masking and DDL. This work also provides detailed information on the implementation of key components such as memory protection unit, clock edge randomizer, and random number generator (RNG).
 
We fabricated a testchip in 65 nm CMOS. We report the area cost of all counter-measures used, as well as the power consumption of each microprocessor and the RNG. The quality of random numbers is evaluated using NIST tests, autocorrelation, and Shannon entropy. Side-channel leakage assessment is performed using an extensive database of recorded power traces---total of more than 40 M traces. We use state-of-the-art analysis algorithms for correlation power analysis (CPA) with dynamic time warping (DTW) pre-processing. Our results evaluate each counter-measure individually, such that we can determine its relative effectiveness.

Our main contributions include the demonstration of a microprocessor where all values are protected with Boolean masking, and precharge logic. This work serves as proof of concept for a system where software implementations can run with fewer counter-measures. Our RTL design and implementation scripts are publicly available~\cite{resCodeSCA}. We describe scalable, and automated solutions that require no changes to the architecture being protected. Relevant problems such as the high-throughput requirements for the RNG, clock randomization, and the details of address/data protection at memory interfaces are discussed. Finally, the paper also performed security assessment of different counter measures using distinct implementations of the same initial design.

\section{Related Works \label{sec:relworks}}

\begin{figure}[t]
    \centering
    \includegraphics[scale=1]{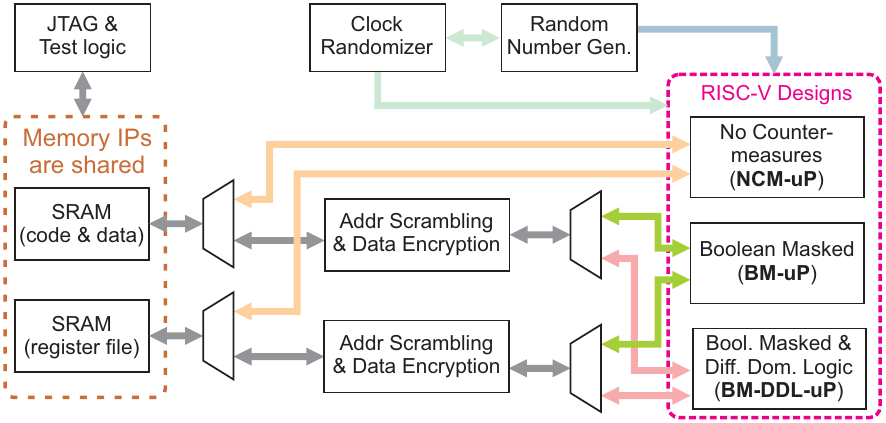}
    \caption{Top level block diagram of our testchip.}
    \label{fig:toplevel}
\end{figure}

Secret keys can be extracted from the power consumption or electromagnetic emanations of unprotected devices~\cite{seminalDpa1999, seminalEMA2001}. Boolean masking was first proposed in~\cite{mskFirst1999}, followed by~\cite{mskTrichina2003, mskProbing2003}. Threshold implementation (TI) was introduced as a provably glitch resistant masking technique~\cite{mskThreshold2006}. Recent developments have been focused on reducing the area and randomness cost of implementing masked circuits~\cite{mskBiru2017, mskConsolidating2015, mskDomain2016, mskFuckBiru2019, mskLowLatency2018, mskReconciling2017, mskUnified2018}. Most of Boolean masking literature is centered at protecting non-linear operations in cryptographic algorithms. Interest in larger scopes of protection has risen in recent years. A microprocessor ALU is masked using TI~\cite{mskThresholdArith2015}. Masking was applied to a binarized neural network~\cite{mskBomanet2020}, and to RISC-V microprocessors~\cite{mskDomainUp2016, mskRambus2019}. This work explores masking expressions that have lower cost and are easy to integrate into already existing designs~\cite{mskBiru2017}.

Differential logic styles with return to zero encoding make power consumption less data-dependent. For example, in~\cite{lsSABL2002} a sense amplifier based logic (SABL) is used to reduce asymmetries of the dynamic logic. In~\cite{lsWDDL2004}, wave dynamic differential logic (WDDL) uses static CMOS gates to replicate the behavior of dynamic logic, but glitches still leak sensitive information. Other logic styles include~\cite{lsDRP2004, lsMDPL2005, lsTDPL2006, lsTEL2018, lsMDPL80512007, lsAdiabatic2014, lsMultitop2014}. Our work differentiates in the applied methodology. Instead of ``semi-custom" techniques, our scripts are integrated with CAD tools and perform automatic replacement of relevant static CMOS cells by dynamic cells, without affecting place and route or static timing analysis.

Often, counter-measures are tailored to the hardware implementation of specific algorithms. For example, in~\cite{aesDualrail2020}, authors use randomized byte-ordering, heterogeneous substitution boxes, and linear Boolean masking of mix-columns to protect an AES implementation. Other recent works have developed more generic protection mechanisms using fast, randomized, voltage dithering in the voltage regulator~\cite{aesDither2018, aesDCDC2021, aes4K2021, aesStellar2021}. Comparing the effectiveness of counter-measures from different publications is not trivial. Changes in the acquisition system, and the presence of combined counter-measures make it hard to draw any conclusions. Finally, it is possible that small drops in the supply voltage, could create misalignment in the power traces. The mentioned works lack a side-channel leakage assessment using alignment techniques such as dynamic time warping (DTW).

High-throughput RNGs are essential for effective masking of digital circuits. Large literature exists for harvesting random numbers from different entropy sources. Recent proposals include latch metastability~\cite{rngLatchIntel2016, rngLatchJapan2022, rngSRAM2018}, time to collapse in multi-mode ring-oscillators~\cite{rngThreeEdge2016}, or common mode analog comparators~\cite{rngCommonMode2016}. But perhaps the most well understood entropy source is still phase jitter~\cite{rngModel2011}. While cryptographic algorithms need high-quality random numbers with forward/backward secrecy~\cite{stdAIS312011}, masking implementations may tolerate random numbers generated by lightweight RNGs designs that focus on throughput---and that is where our RNG is positioned.

\section{Hardware Architecture \label{sec:hwarch}}

\subsection{Top Level Design}

Our design has three RISC-V microprocessors. See Fig. \ref{fig:toplevel}. Each microprocessor is designed to test a different set of security counter-measures. They share two SRAM memories for code, data, and register file (RF). The test logic selects which one of the three microprocessor will be operational, as well as the desired configuration for clock edge randomization and random numbers. The RISC-V implementation with no counter-measures (NCM-uP) has direct access to the SRAM memories, while others go through a memory protection unit. Netlist manipulation techniques described in this section are only applied to the BM-uP and BM-DDL-uP microprocessors. Other blocks are synthesized with the typical ASIC digital design flow, using a commercial library of standard cells.

\subsection{Bit-serial RISCV Microprocessor}

\begin{figure}[t]
    \centering
    \includegraphics[scale=1]{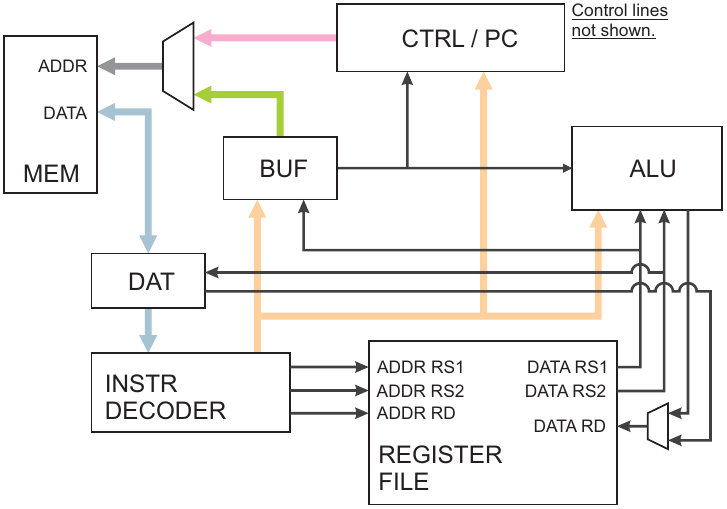}
    \caption{Bit-serial RISC-V datapath. Control lines are not shown. Thin lines have width of a single bit. Although not shown in this diagram, memory and register file are shared among the three microprocessors.}
    \label{fig:serv}
\end{figure}

We used a bit-serial CPU, show in Fig \ref{fig:serv}. It is based on a publicly available design\footnote{https://github.com/olofk/serv}. This CPU design was chosen for its small area footprint. The internal datapath is one bit wide. Techniques described in this work are equally applicable to larger microprocessors.  Main memory is single port and uses 32 bit words to store both code and data. The register file (RF) is dual port. One port for writes, another for reads with 2 bit words to avoid an additional read port. Arithmetic logic unit (ALU) operates on a single bit of data per cycle. The fetch-execute-writeback process requires 36 cycles. Instructions need one, or two phases to complete. Two-phase instructions such as loads, stores, and branches can use up to 70 cycles. Only three 32 bit registers exist. BUF is a 32 bit register that holds data between phases, which can be addresses for branches, load/store, or data to be shifted. DAT is another 32 bit register, used for memory load and store operations, where data is shifted in from RS2 during store operations, and shifted out to RD during load operations. The last 32 bit register holds the program counter (PC). If the instruction is not a branch, next address (PC + 4) is calculated one bit at a time, in parallel with ALU execution. All three 32 bit registers work similarly to a shift register during execution. For memory operations, parallel output/capture functionality is used. The instruction decoder (ID) parses instructions from DAT. It outputs immediate fields to BUF, CTRL/PC, and ALU, one bit at a time, to calculate jumps addresses and arithmetic operations. Control and status registers (CSRs), interrupts, multiplication, and divide instructions were not implemented. Using 65 MHz clock, an AES encryption of 128 bits, without any software counter-measures, takes nearly 20 ms---substitution boxes (SBOXes) not stored as lookup tables, but computed during execution.

\subsection{Clock Generation and Randomization}

\begin{figure}[t]
    \centering
    \includegraphics[scale=1]{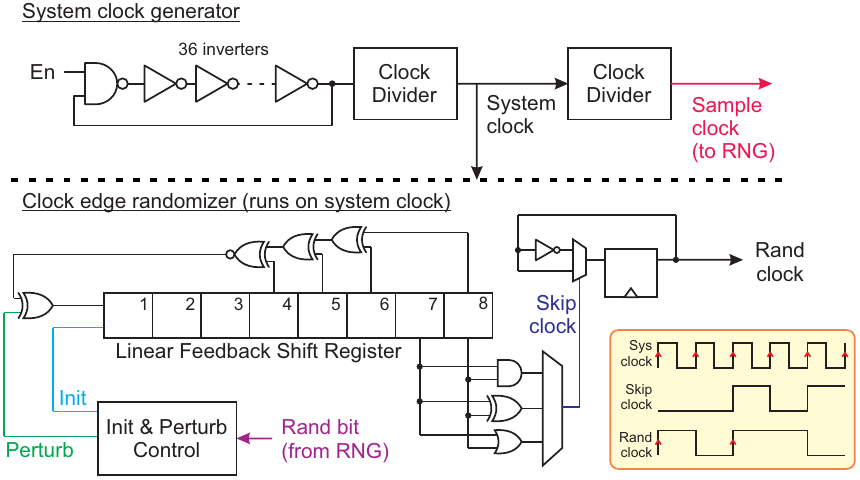}
    \caption{Clock generation and randomization logic. Glitch suppression, test, and forbidden state prevention logic are not shown.}
    \label{fig:clkrnd}
\end{figure}

Power analysis attacks require a large number of power traces. Each trace records the power consumption while the device is manipulating sensitive information with different input data. For example, AES encryptions with different plain-text, but same key. As discussed in section \ref{sec:meas:cpa}, the effectiveness of power attacks is significantly higher if traces are aligned in time. Therefore, the insertion of random delays to purposely misalign power traces may be used as counter-measure against power analysis attacks. To avoid changing software or existing hardware architectures, we insert random delays by manipulating the clock signal.

Our clock generation and randomization logic is shown in Fig. \ref{fig:clkrnd}. A ring-oscillator (RO) generates the clock using 36 inverters and a NAND gate. Layout of the ring-oscillator was done manually. We used MOSFETs with channel length 4x the minimal value to achieve nominal frequency of 226 MHz. Compensation for temperature and noise was not implemented. The hardware includes an adjustable clock divider from 2 to 256. Glitch suppression and test logic are not shown. To generate a random signal for clock edge randomization we used an 8-degree LFSR in XNOR form, polynom $x^8+x^6+x^5+x^4+1$, cycle length $2^8-1$, all ones excluded~\cite{rngLfsrXilinx1996}. Output is taken from the two least significant bits, it indicates when an edge will be skipped. A multiplexer was added to select how often clock edges are skipped. Valid options are 25\%, 50\%, or 75\% for outputs from AND, XOR, and OR, respectively. For example, if LFSR state is ``10001101" both XOR and OR outputs would skip next clock edge if selected, while the AND output would introduce a clock edge.

\begin{figure}[t]
    \centering
    \includegraphics[scale=1]{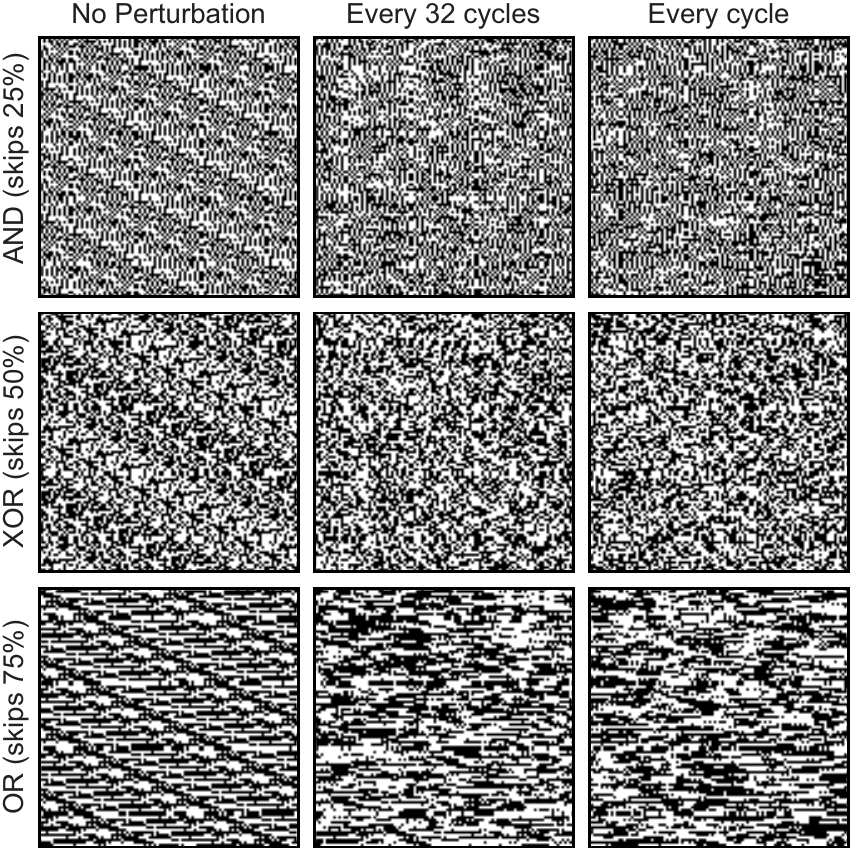}
    \caption{Clock waveform visualization for 9 different randomization settings. White/black pixels represent zero/one. Time passes from left to right, top to bottom. White pixel on the left, and black pixel on the right represents a rising edge.}
    \label{fig:clkimg}
\end{figure}

Real applications require the hardware to complete operations under a time budged. We generate the clock randomization pattern using an 8~bit LFSR, which has sequence length of 255. Therefore, within a 255 cycles window, the randomized clock will have a constant number of edges. This enables performance predictability with resolution of 255 system clock cycles. Nevertheless, one may argue that a repeating clock randomization pattern introduces potential vulnerabilities. For this reason, we allow the 8~bit LFSR to be perturbed, at an user adjustable rate. The perturbation signal comes from the RNG, therefore, it has very large cycle length. If perturbed every clock cycle, we can assume that the randomized clock pattern will never repeat. However, under such frequent perturbation rate, the LFSR is more likely to, sometimes, produce a run of 30 ones---causing the clock randomizer to skip 30 consecutive clock edges. To reduce the likelihood of such events, we suggest a moderate perturbation rate. For example, when perturbed every 32 cycles, the clock randomizer pattern will follow the 8~bit LFSR natural counting sequence for 31 cycles, until it is perturbed again.

Fig. \ref{fig:clkimg} shows a grid of 3 by 3 subplots with randomized clock waveform presented using image pixels. Rows have different clock skip rates, while columns have different perturbation rates. In each subplot, time is discretized as pixels. White/black pixels represent logic 0/1, respectively. Time passes from left to right, top to bottom (the clock waveform is folded in the image area). A transition from white pixel (at the left) to black pixel (at the right) denotes a rising edge. A pattern is clearly visible in the clock waveform when the LFSR is not perturbed. Perturbations every 32 cycles, and every cycle, remove noticeable patterns.

\begin{figure}[t]
    \centering
    \includegraphics[scale=1]{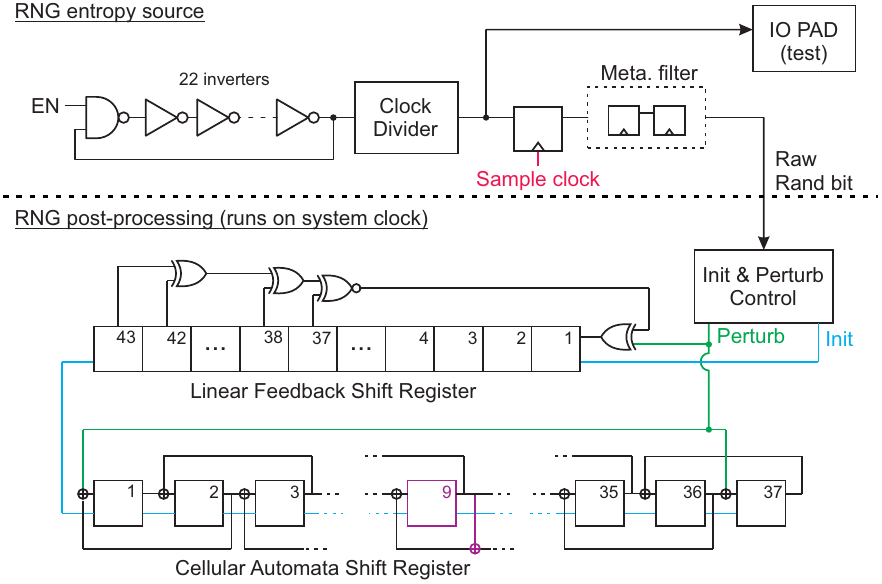}
    \caption{Random number generator architecture. The sampling clock comes from the clock generation circuit.}
    \label{fig:rng}
\end{figure}

The clock skip rate affects the number of rising edges occurring in a period of time. Application engineers may assign different clock skip rates to software routines, depending on the sensitivity of data being manipulated and performance requirements. Also, care must be taken to avoid placing the LFSR into forbidden state (``all ones", where updates no longer occur). The LFSR update cycle is skipped in case the next state is all ones. The initial state is generated by the RNG and loaded serially during power-on.

\subsection{Random Number Generator}

Random numbers are a critical component of secure ICs. Both clock randomization and Boolean masking require random numbers to work effectively. In particular, Boolean masking needs a large number of fresh (new) random numbers every clock cycle for remasking operations (see section \ref{sec:hwarch:mask}). The entropy requirement for Boolean masking however, is not as high as typical cryptographic uses such as key generation. Therefore, our random number generator (RNG) focus on high throughput to attend the demand of our implemented counter-measures.

Our RNG design uses ring-oscillator (RO) thermal noise as entropy source. It accumulates thermal noise over a period of time, which produces phase jitter. The entropy of a sampled value is inversely proportional to the sampling rate. The RNG RO design, shown in Fig. \ref{fig:rng}, uses same techniques as the system clock RO, but with relatively prime number of stages to avoid frequency locking. The RNG RO has 22 inverters and a NAND gate, with 366 MHz nominal frequency. The sampling clock is derived from system clock. Sampling frequency depends on target entropy rate and on technology noise characteristics. An IO PAD is connected to the RO output for noise characterization (see section \ref{sec:meas:rng}). A two stages meta-stability filter running on system clock was added to the raw random bit output.

\begin{figure}[t]
    \centering
    \includegraphics[scale=1]{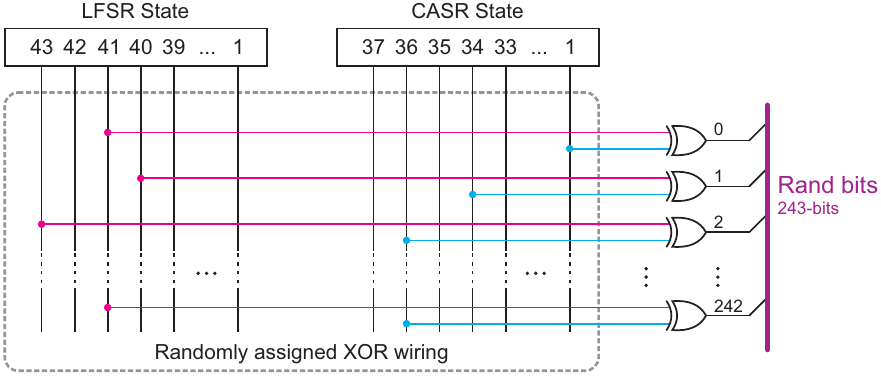}
    \caption{Output XOR network for random numbers. XOR wiring is defined randomly at design phase.}
    \label{fig:rngwires}
\end{figure}

\begin{figure}[t]
    \centering
    \includegraphics[scale=1]{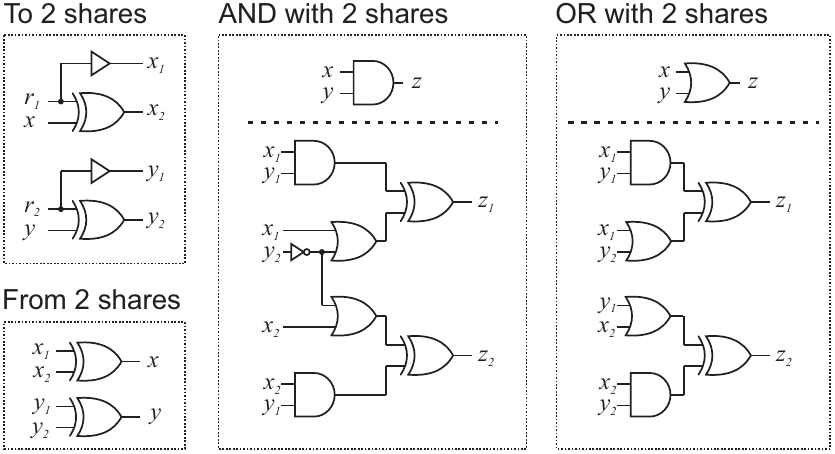}
    \caption{Boolean masking expressions to perform non-linear operations on shared values.}
    \label{fig:boolmask}
\end{figure}

Throughput of raw random bits depends on sampling clock frequency, which is typically much lower than system clock frequency. To increase throughput and remove possible entropy source biases, we added a post-processing block. Our solution is based on~\cite{rngMotorola2002}. We use an 43-degree LFSR in XNOR form, polynom $x^{43} + x^{42} + x^{38} + x^{37} + 1$, cycle length $2^{43}-1$, all ones excluded~\cite{rngLfsrXilinx1996}. We also used a one-dimensional linear hybrid cellular automata shift register (CASR) of 37 cells and cycle length of $2^{37}-1$, all zeros excluded~\cite{rngHyridCa1995}. The hybrid CASR uses update rule 90 in all states, except in position 9, where rule 150 is used. Rule 90 and rule 150 are defined as $a_i(t + 1) = a_{i-1}(t) \oplus a_{i+1}(t)$, and $a_i(t + 1) = a_{i-1}(t) \oplus a_i(t) \oplus a_{i+1}(t)$, respectively~\cite{rngWolframCa1986}. The LFSR/CASR states are initialized with raw random numbers during power on. Since the cycle length of LFSR and CASR are relatively prime, the combined cycle length is close to $2^{80}$. Both LFSR and CASR update at every system clock cycle.

LFSR feedback and the CASR first/last cells are XORed with the new raw random bit prior to update. These perturbations to the natural counting sequence only occur when a new raw random bit is available, which depends on the sample clock frequency. Output is derived by XORing LFSR and CASR states. We use an XOR network, shown in Fig. \ref{fig:rngwires}, that derives 243 outputs from LFSR/CASR state. Up to 1591 outputs are possible using two input XOR gates. More outputs require the XORing more than two state bits. The connection pairs are unique, randomly assigned, and defined at design time (they are static).

\subsection{Boolean Masking \label{sec:hwarch:mask}}

\begin{figure}[t]
    \centering
    \includegraphics[scale=1]{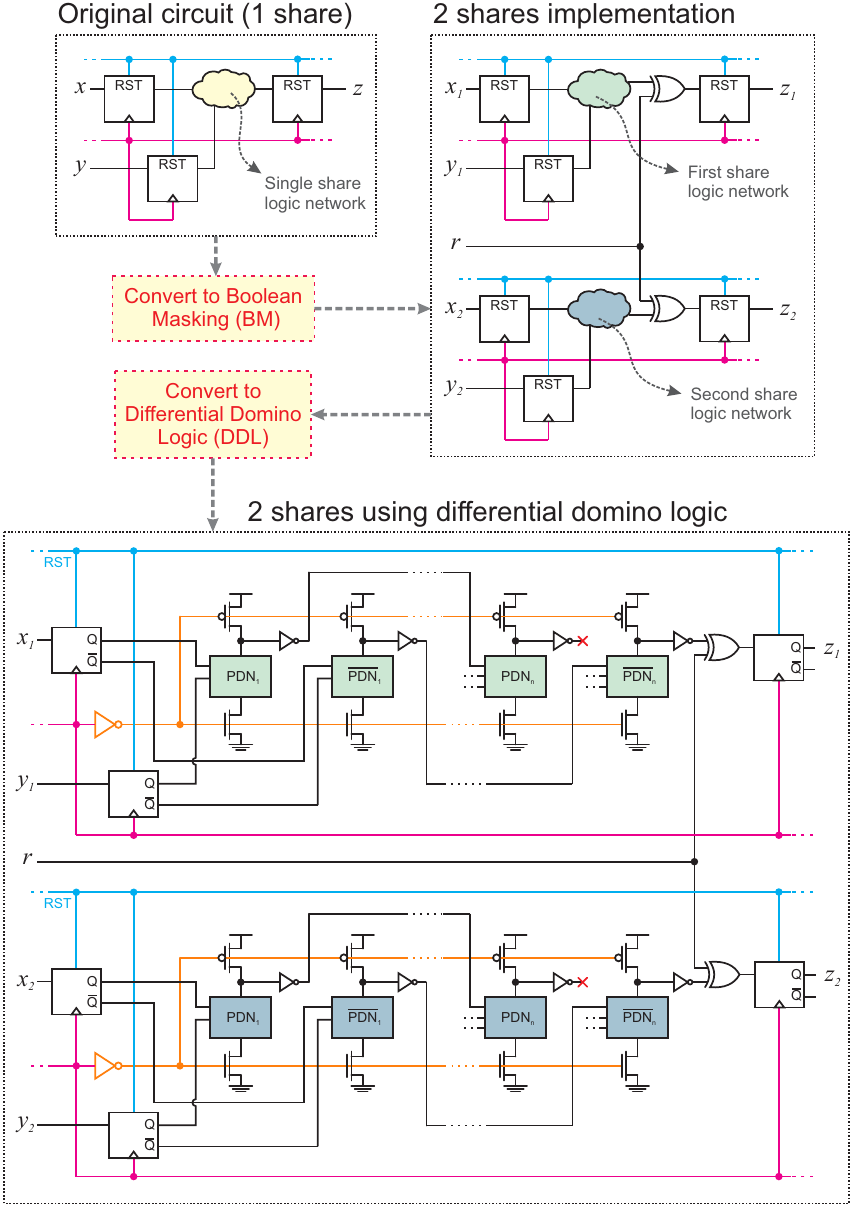}
    \caption{Netlist conversion from original circuit, to shared, and then dynamic logic.}
    \label{fig:onetwoshares}
\end{figure}

Boolean masking (BM) splits each value into two or more shares that are (ideally) uncorrelated to the original value. Computation using multiple shares requires careful logic manipulation, where each operation is replaced by its masked counterpart. We applied Boolean masking to the microprocessors designated as BM-uP and BM-DDL-uP. We used masking expressions from~\cite{mskBiru2017}. Although the masking expressions themselves are secure, further linear combinations can make them insecure~\cite{mskFuckBiru2019}. Nevertheless, the used expressions have key characteristics that are very attractive for the construction of \textit{fully-masked} large-scale designs: i) they are compact, and ii) do not require fresh randomness at every operation. Fig. \ref{fig:boolmask} shows the masked implementation of basic non-linear logic gates (AND, OR) using masking expressions from~\cite{mskBiru2017}. Linear operations such as XOR, shifts, and permutations do not require any modification, they are applied to both shares equally. Fresh random numbers are required when converting from single share to two shares. Conversions between 2 shares and single share are only performed in the memory protection module, while data is still encrypted by session keys. Therefore, plain-text values are never exposed.

We developed a script that converts arbitrary circuit netlists into their masked counterparts. The script is integrated with CAD tools. Two restrictions were imposed: i) the input netlist can not use clock and reset as data; and ii) reset must be asynchronous. See Fig. \ref{fig:onetwoshares}. The script splits every signal, except reset and clock, in two shares, replacing all basic operations by their masked counterparts. Circuit functionality and latency are not affected. All registers are duplicated and wired to accommodate the additional share of each state. An XOR gate is added before every register to refresh the masking with a new random number every cycle. No random number is ever reused. New top level ports are automatically created to accommodate the additional share of all previously existing input/output ports. A new top level input port bus for fresh random numbers is also created. It is driven by the RNG output network, shown in Fig. \ref{fig:rngwires}, and each bit connects directly to a pair or remasking XOR gates.

\begin{figure}[t]
    \centering
    \includegraphics[scale=1]{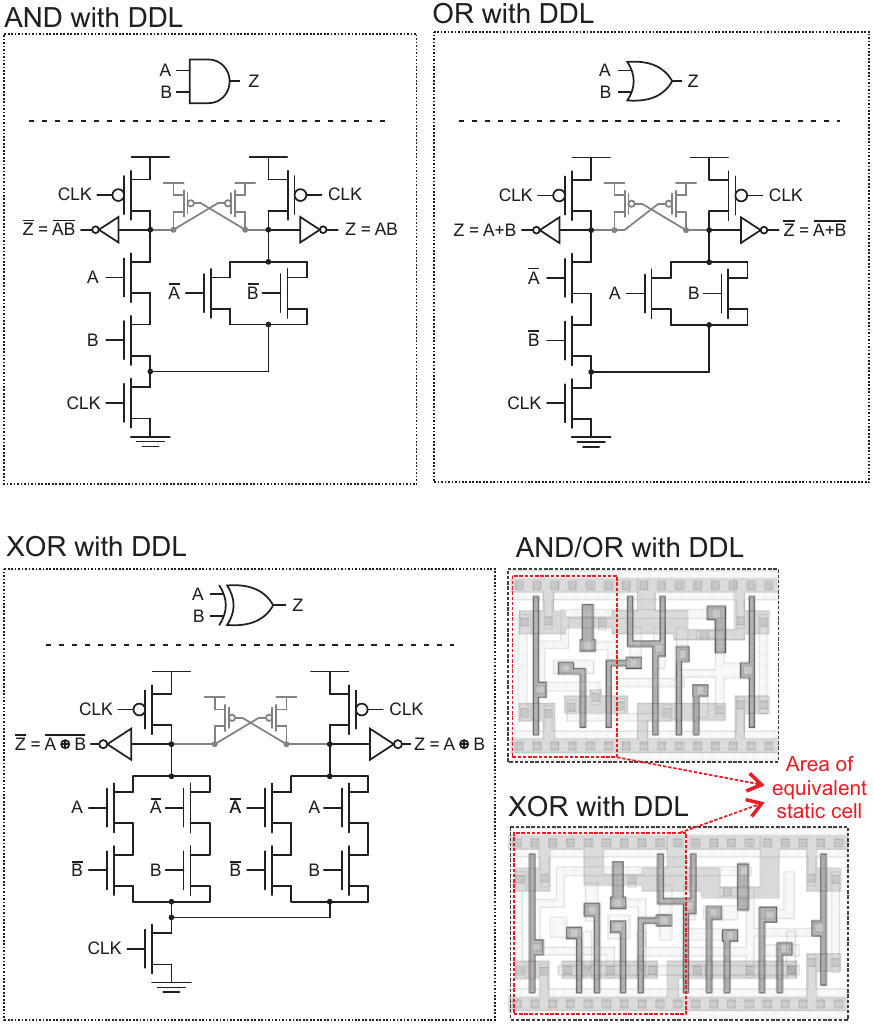}
    \caption{Dynamic domino gates and associated layout.}
    \label{fig:ddl}
\end{figure}

The masked implementation of non-linear operations requires signals to traverse between first/second logic networks. Such signals are not represented in Fig. \ref{fig:onetwoshares}. Also, our implementation does not mask the reset signal. If reset is synchronous, it can be masked together with logic.

\subsection{Differential Domino Logic}

\begin{figure}[t]
    \centering
    \includegraphics[scale=1]{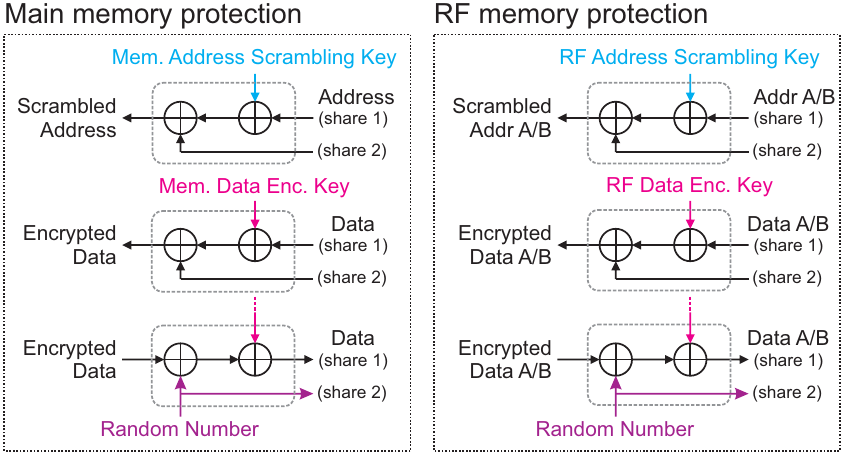}
    \caption{Memory protection scheme.}
    \label{fig:memprot}
\end{figure}

Masking expressions for non-linear operations (AND, OR gates) use signals from both shares (see Fig. \ref{fig:boolmask}). Glitches in that logic may momentarily expose protected values, reducing the effectiveness of Boolean masking~\cite{mskGlitches2005}. Traditional solutions use registers to stop glitch propagation, which modifies circuit latency and requires significant design effort for existing designs. We reduce glitches by implementing our logic gates using differential domino logic (DDL), a differential input/output dynamic logic style with precharge/evaluation phases ~\cite{bookRabaey2002}. During precharge, both outputs are driven to the same value, but only one of them toggles in the evaluation phase. Logic gate inputs are restricted to a single 0 to 1 transition. Fig. \ref{fig:ddl} shows the transistor level schematic of AND, OR, and XOR using DDL, with associated layout. Inverters are implemented by swapping the output wires. AND/OR gates have the same layout, which could be convenient for circuit obfuscation techniques, such as split foundry manufacturing~\cite{obfSplit2013}.

Our netlist conversion script was adapted to use our DDL logic gates instead of the typical CMOS library cells. The output netlist has each wire split in two shares for Boolean masking. Moreover, each share is again split in two complementary signals for DDL. This technique was applied to BM-DDL-uP, in Fig. \ref{fig:toplevel}. The layout of our DDL cells has same height of the CMOS library cells, so that we can conveniently mix DDL gates with static CMOS gates from the commercial library. Better area results are possible with taller DDL cells, but interoperability with commercial library cells would be extremely hard.

Our output netlist uses sequential elements from the commercial library. The remasking XOR gate preceding every sequential element also uses a cell from the commercial library---its output has no connection to dynamic cells, and glitches in that gate are unlikely to cause significant information leakage since one of the inputs is a random number. Inverted inputs to DDL gates are wired to registers inverted output pins. Complementary signals were not granted their own sequential element---the number of registers do not quadruple with respect to the original design. Inverted outputs from the last DDL gate in a combinational path (prior to the remasking XOR) are left unconnected. This is a source of load imbalance that will leak information, but it avoids significant area cost.

We characterized multiple strengths of the DDL cells using a commercial tool. Our custom DDL library has two strengths of the AND/OR gate, and three strengths of the XOR gate. Dynamic logic is not supported by the characterization tool, so we manually specified the relevant input/output timing arcs for delay characterization. We also wrote Verilog models for functional simulation, supporting delay annotation. The Liberty and Verilog files were used for timing analysis and design sign-off.

In order to maintain the methodology applicable to complex digital designs, we did not use routing constraints for differential signals. Therefore, wiring capacitance imbalances may be a source of information leakage.

\subsection{Memory Protection \label{sec:hwarch:memprot}}

\begin{table}
    \centering
    \caption{Area utilization in 65 nm CMOS.}
    \label{tab:area}
    \begin{threeparttable}
        \begin{tabular}{@{}lrr@{}}
\toprule
\textbf{Unit name}                    & \textbf{\# of Instances}                         & \textbf{Area ($\mu m^2$)}               \\ \midrule
Clock generation \& randomization     & 203                                              & 922                                           \\
Random number generation              & 674                                              & 2735                                          \\
Microprocessor NCM-uP                 & 731                                              & 3509                                          \\
Microprocessor BM-uP                  & 8232                                             & 22967                                         \\
Microprocessor BM-DDL-uP              & 7633                                             & 44426                                         \\
Main memory (code \& data)            & 13                                               & 39505                                         \\
Register file                         & 19                                               & 10720                                         \\
Memory protection and muxes           & 768                                              & 2653                                          \\ \bottomrule
\end{tabular}
        \begin{tablenotes}[para,flushleft]
            Notes: XOR output network uses 242 XOR instances and account for 871-$\mu m^2$ of the area reported for the RNG.
        \end{tablenotes}
    \end{threeparttable}
\end{table}

\begin{figure}[t]
    \centering
    \includegraphics[scale=1]{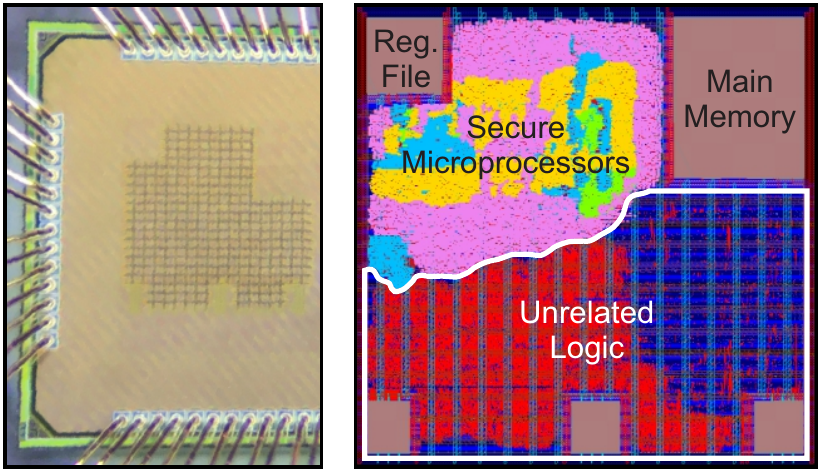}
    \caption{Die photo of the implemented chip in 65 nm CMOS technology and its layout. The layout highlights standard cells for NCM-uP (green), BM-up (yellow), and BM-DDL-uP (pink). Test logic, RNG, clock generation/randomization are shown in light blue.}
    \label{fig:chip}
\end{figure}

A trivial solution for memory protection is to duplicate the memory size and store both shares of the Boolean masked data. To avoid a twofold increase in the SRAM size, we perform encryption using a session based key, which is XORed with the data to alter its hamming weight. Moreover, we also scramble the address bus using another session based key that is XORed with the address, changing the memory address layout. Session keys have different values at every session. The definition of a session is application dependent, but in our case, it corresponds to one execution of the target program in one of the microprocessors. Session keys remain stable for the duration of a session. Our session keys are provided via test logic. Fig. \ref{fig:memprot} shows the memory protection blocks. The order in which the XOR operations are performed is extremely important to avoid exposing plain-text values.

Single-port memory protection can be improved further. The encrypted data can be XORed with the address before it is written, making data encryption address dependent. This requires a non-linear expansion of the 10 bit address into a 32 bit word, which has extra area costs. It is also important to notice that our memory protection unit does not include firewall capabilities to detect unauthorized accesses. If present, such logic must be protected with Boolean masking and DDL.

\section{Measurement Results \label{sec:meas}}

\subsection{Chip Area Utilization}

\begin{figure}[t]
    \centering
    \includegraphics[scale=1]{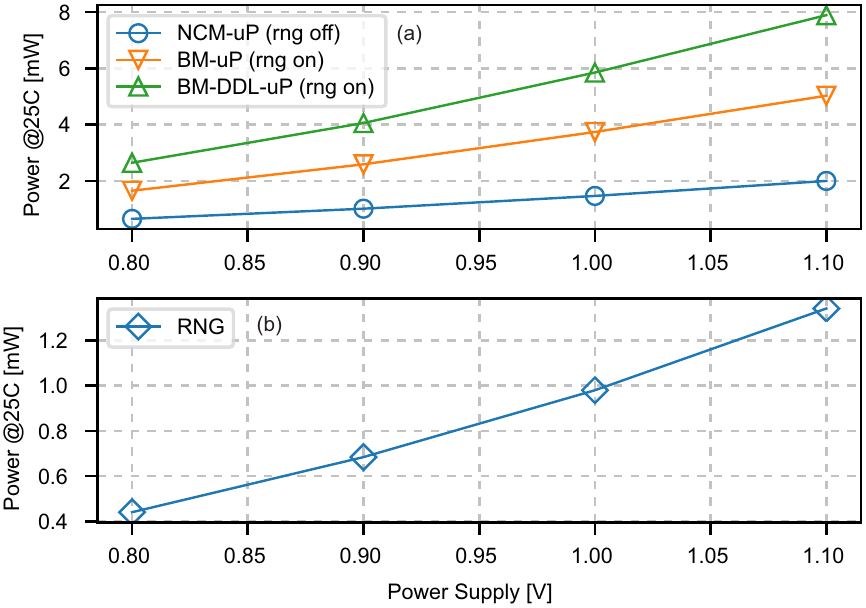}
    \caption{Power consumption of (a) all RISC-V microprocessors, (b) RNG. Static components are not included.}
    \label{fig:power}
\end{figure}

\begin{table}
    \centering
    \caption{NIST tests on RNG output.}
    \label{tab:sts}
    \begin{threeparttable}
        \begin{tabular}{@{}lrr@{}}
\toprule
\textbf{Test Name}      & \multicolumn{1}{l}{$\rho$} & \textbf{Proportion} \\ \midrule
Frequency               & 0.554420                       & 98/100              \\
Block Frequency          & 0.042808                       & 98/100              \\
Cumulative Sums          & 0.964295                       & 97/100              \\
Runs                    & 0.013569                       & 99/100              \\
Longest Run              & 0.289667                       & 98/100              \\
Rank                    & 0.249284                       & 98/100              \\
FFT                     & 0.474986                       & 99/100              \\
Non Overlapping Template  & 0.867692                       & 96/100              \\
Overlapping Template     & 0.115387                       & 100/100             \\
Universal               & 0.334538                       & 97/100              \\
Approximate Entropy      & 0.181557                       & 99/100              \\
Random Excursions        & 0.941144                       & 67/69               \\
Random Excursions Variant & 0.619772                       & 66/69               \\
Serial                  & 0.224821                       & 98/100              \\
Linear Complexity        & 0.017912                       & 100/100             \\ \bottomrule
\end{tabular}
        \begin{tablenotes}[para,flushleft]
            Notes: Cumulative Sums, Non Overlapping Template, Random Excursions, Random Excursions Variant, and Serial run multiple times in the NIST SP 800-22 Rev1a. Values in table refers to the worst pass proportion among all runs.
        \end{tablenotes}
    \end{threeparttable}
\end{table}

We fabricated a testchip in a 65 nm CMOS process, shown in Fig. \ref{fig:chip}. Table \ref{tab:area} shows the number of instances and area utilization of each block after design sign-off. The clock generation includes a 57-$\mu m^2$ RO and all clock edge randomization logic. The RNG includes a 37-$\mu m^2$ RO and all post-processing logic (LFSR, CASR, and XOR network). The XOR network uses 242 XOR instances and accounts for 871-$\mu m^2$ of the reported area for RNG. The main memory uses a single-port 32 kbit SRAM (1024x32). RF uses a dual-port 1 kbit SRAM (512x2). The BM-uP implementation had an area increase of 6.5x compared to baseline (NCM-uP), while the number of instances increased by 11.3x. The BM-DDL-uP had an area increase of 1.9x compared to BM-uP, but the number of instances showed small reduction due to the absence of inverters in differential logic styles.

\subsection{Power Consumption}

Dynamic power consumption was measured from 0.8V up to 1.1V, for the NCM-uP, BM-uP, BM-DDL-uP, and RNG. Nominal voltage is 1.0 V (see Fig. \ref{fig:power} (a) and (b)). We share the same power supply pin with other non-related digital circuits which had their clock disabled during experiments. Moreover, static power was measured and excluded from all reported values. The RNG was not active during measurements of NCM-uP. The power reported for the RNG includes system clock oscillator for sampling, RNG oscillator, LFSR/CASR post-processing, XOR output network, and remasking XOR gates.

\begin{figure}[t]
    \centering
    \includegraphics[scale=1]{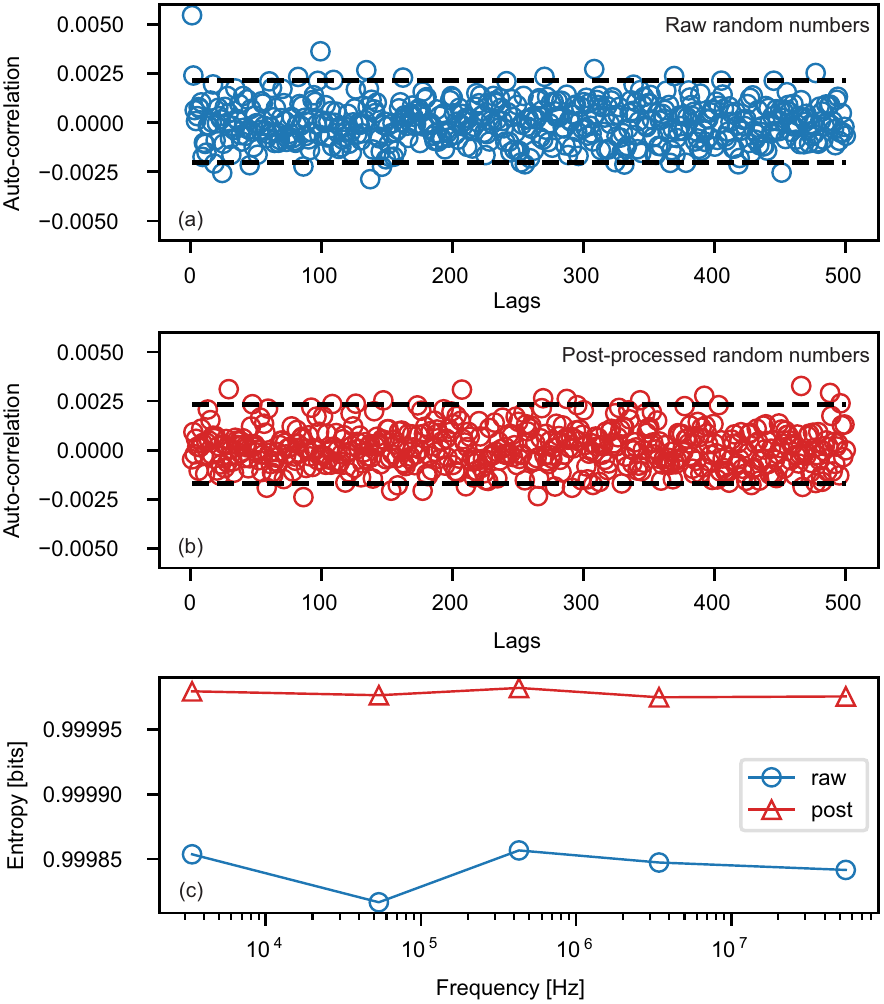}
    \caption{Auto-correlation tests on (a) raw, and (b) post-processed random numbers. Shannon entropy tests in (c).}
    \label{fig:rngtest}
\end{figure}

\subsection{Quality of Random Numbers \label{sec:meas:rng}}

The National Institute of Standards and Technology (NIST) published tests to assess the quality of random numbers~\cite{stdNIST2010}. Table \ref{tab:sts} reports results for all tests. A total of 100 bitstreams with 1 M bits each was used. All tests meet the suggested passing criterion of 80\%. Fig. \ref{fig:rngtest} plots the autocorrelation of 1 M (a) raw random bits, and (b) after post-processing by LFSR/CASR logic. We used a sampling frequency of 3.4 MHz. Dashed lines represent the interval that contains 95\% of the data. Fig. \ref{fig:rngtest} (c) plots the Shannon entropy for 1 M bits at various sampling frequencies. Entropy was calculated using intervals of 5 bits.

An statistical modeling of phase jitter entropy sources was presented in~\cite{rngModel2011}. Authors derived an expression for minimum entropy rate. To use their expression, we measured the clock period standard deviation of our RNG oscillator over 120 k cycles. Our testchip divides the clock by 256 and routes the signal to an output pin. Measured clock period mean was 716 ns, with 251 ps of standard deviation, which reflects the accumulated thermal noise in our RNG oscillator. Oscillators with high standard deviation values harvest more entropy for the same sampling period. Similar jitter is also present in our system clock oscillator and it was taken into account during timing analysis as clock uncertainty---it does not affect circuit functionality. The calculated minimum entropy for raw random numbers sampled at 3.4 MHz is 0.41 bits (per random bit). Using the model, 1.0 bit of entropy is achieved with sampling frequency of 200 Hz or lower. Nevertheless, our application tolerates lower entropy for increased throughput. If a higher minimum entropy at increased throughput is required, additional ROs can be XORed before the sampling register. In that case, it is imperative to use ROs with relatively prime number of stages to avoid frequency locking. Also notice that the model does not include phase jitter from our sampling clock, which could increase the calculated minimum entropy.

\begin{table}[t]
    \centering
    \caption{Comparison of RNG results.}
    \label{tab:rngcmp}
    \begin{threeparttable}
        \begin{tabular}{@{}lccccc@{}}
\toprule
                & \textbf{This work}    & \textbf{JSSC'16}       & \textbf{JSSC'16}       & \textbf{JSSC'22}       \\
&  & \cite{rngLatchIntel2016} & \cite{rngCommonMode2016} & \cite{rngLatchJapan2022} \\ \midrule
Technology      & 65~nm        & 14~nm         & 65~nm         & 130~nm        \\
Entropy         & Jitter       & Meta          & Meta+Jitter   & Meta          \\
Bit rate (Gb/s) & 12.8         & 0.225         & 3             & 0.002         \\
Area ($\mu m^2$)& 2735         & 1088          & 1609          & 5561          \\
Power (mW)      & 0.992        & 1.5           & 5             & -             \\
Energy (fJ/bit) & 0.078        & 6.67          & 1.67          & -             \\
Post-processing & LFSR/CASR    & AES           & None          & VN8W          \\ \bottomrule
\end{tabular}
        \begin{tablenotes}[para,flushleft]
            Notes: Power and energy reported for our work do not include the static component. 
        \end{tablenotes}
    \end{threeparttable}
\end{table}

Our RNG produces 243 random bits each clock cycle, therefore its total throughput at 56 MHz is 12.8 Gb/s. Further details such as area, power, and energy efficiency are provided in Table \ref{tab:rngcmp}. This lightweight, throughput focused RNG design meets our requirement for Boolean masking operations, but arguably, other sensitive contexts such as key generation, padding, and nonces will require minimum entropy guarantees, slower sampling frequencies, and forward/backward secrecy in the post-processing---which increases the hardware size significantly. We recommend the interested reader to see AIS 31 standard for details~\cite{stdAIS312011}.

\subsection{Experimental Setup for Side-channel Analysis}

\begin{figure}[t]
    \centering
    \includegraphics[scale=1]{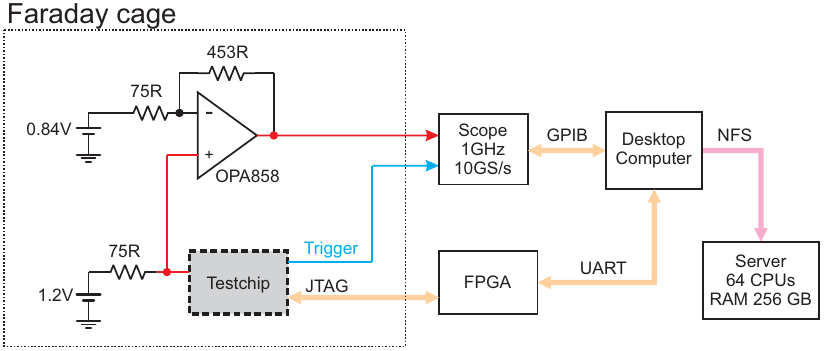}
    \caption{Side-channel analysis acquisition setup.}
    \label{fig:scasetup}
\end{figure}

Fig. \ref{fig:scasetup} shows our experiment setup for side-channel analysis. We added a 75 Ohms resistor in series with the testchip power supply. The signal is conditioned by a high-speed amplifier (OPA858, 5.5 GHz GBW) in a non-inverting configuration with gain of 7. A 0.84 V reference was connected to the gain resistor to avoid clamping of the output signal. The external chip power supply was set to 1.2 V instead of the nominal 1.0 V, to account for the voltage drop in the series resistor. Our oscilloscope has input bandwidth of 1 GHz. The acquisition system is orchestrated by a desktop computer which controls the oscilloscope using GPIB and communicates to the testchip using an FPGA. The FPGA writes the encrypted software using JTAG. All power traces use unique session keys for memory protection. The having sampling rate is 10 GS/s. We used an open source Julia toolbox\footnote{https://github.com/Riscure/Jlsca} to perform the correlation power analysis (CPA) with, and without dynamic time warping (DTW) pre-processing. We attack only the first key byte during an AES encryption operation, targeting the SBOX output using a hamming weight leakage model. Our AES implementation uses multiplicative inverse over Galois field to calculate the SBOX output (lookup tables are not used). Oscilloscope trigger is obtained from a testchip output set by the microprocessor software 36 cycles before the sensitive instruction. The number of samples logged in each power trace is adjusted depending on clock edge randomization settings to ensure all relevant clock cycles are captured.

\subsection{Correlation Power Analysis \label{sec:meas:cpa}}

Correlation power analysis (CPA) uses a set of power traces to extract secret information from a device. Traces record the device manipulating the secret value with different input data. For example, recorded power traces may include many AES encryptions using the same key but different plain texts. CPA attacks extract a few key bits in each iteration. The attacker chooses a key candidate and creates a power consumption hypothesis for each input data, using knowledge of the implemented algorithm, and a leakage model (typically hamming-weight). The correlation between the hypothesis vector and the recorded power traces will be the highest when the correct key is used.

\begin{figure*}[t]
    \centering
    \includegraphics[scale=1]{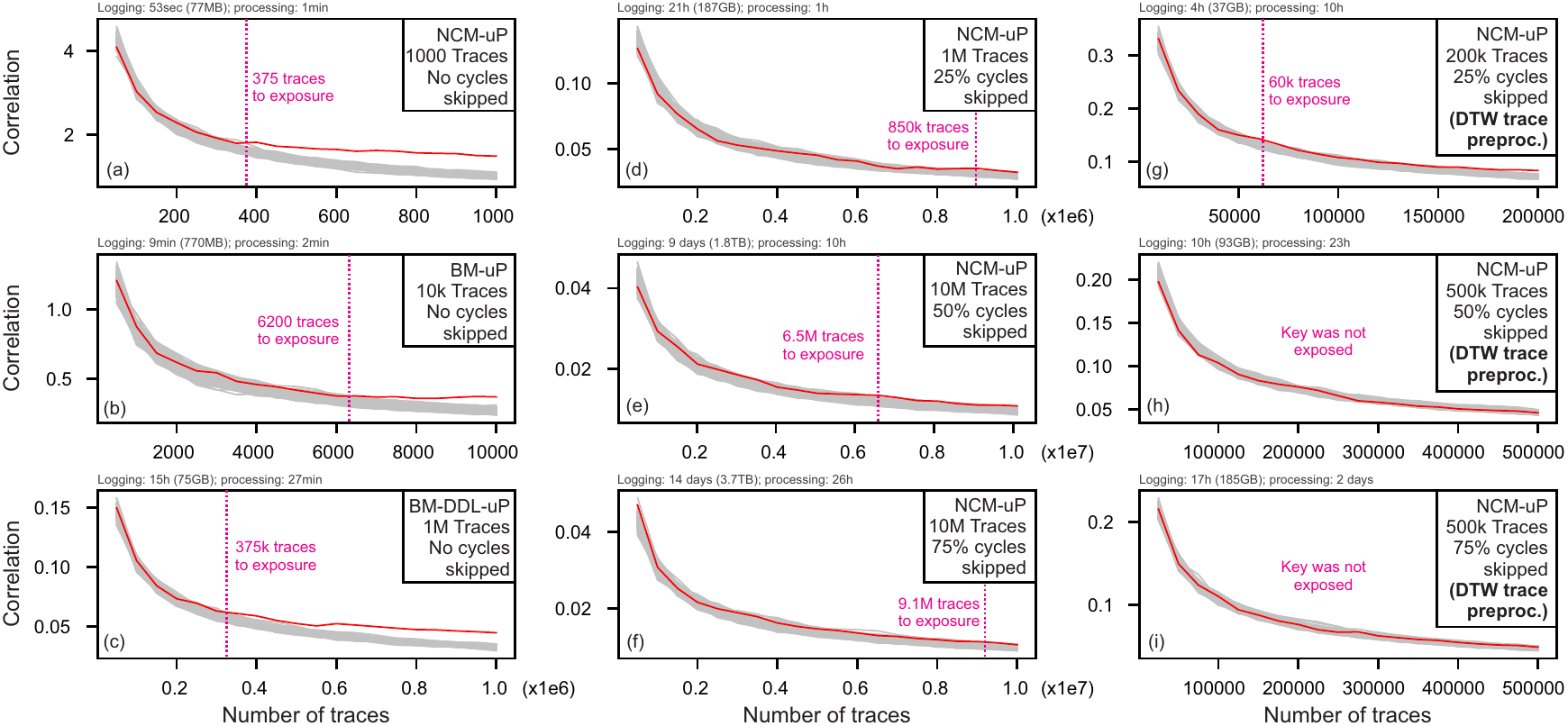}
    \caption{CPA coefficient versus number of traces for different sets of counter-measures. Results without clock edge randomization for (a) NCM-uP; (b) BM-uP, and (c) BM-DDL-uP; results for original netlist (NCM-uP) using clock edge randomization of (d) 25\%, (e) 50\%, and (f) 75\%; and results using DTW pre-processing for trace alignment on NCM-uP with clock edge randomization of (g) 25\%, (h) 50\%, and (i) 75\%.}
    \label{fig:scatable}
\end{figure*}

\begin{figure}[t]
    \centering
    \includegraphics[scale=1]{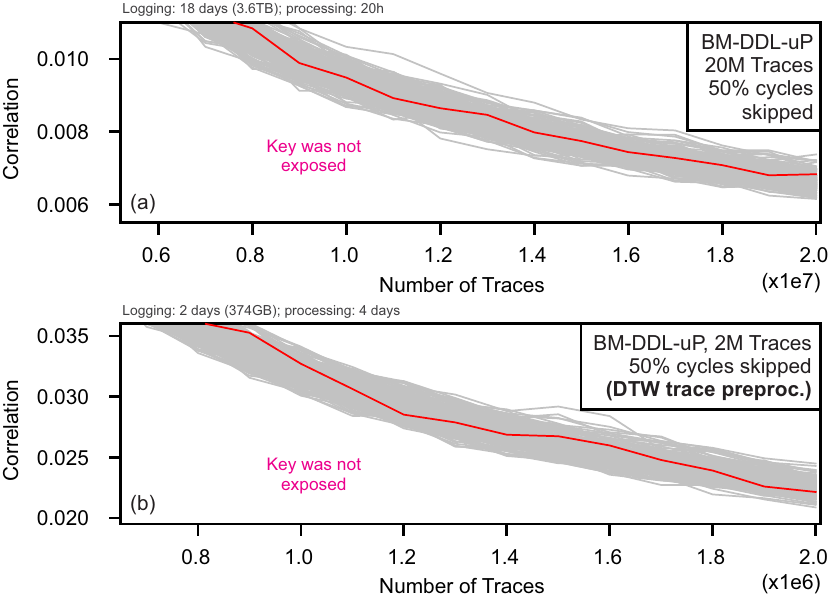}
    \caption{CPA coefficient versus number of traces for BM-DDL-uP with clock edge randomization of 50\%. Results (a) without DTW trace pre-processing and (b) with DTW trace pre-processing.}
    \label{fig:scaall}
\end{figure}

Our side-channel analysis does not include test vector leakage assessment (TVLA)~\cite{stdTVLA2013}. Such tests do not replace conventional key extraction attacks, but provide a quick alternative to detect potential side-channel problems. Nevertheless, they require saving a long power trace that includes a series of encryptions, which would use an enormous amount of oscilloscope memory due to the low throughput of our hardware architecture. Future work shall use a faster architecture so that we can perform these tests.

Fig. \ref{fig:scatable} shows CPA results, plotting the correlation versus number of traces used in the attack. Each plot has 256 lines, one for each key candidate. The correct key (red) is exposed when its correlation is the highest. In Fig. \ref{fig:scatable} (a), the baseline implementation with no counter-measures (NCM-uP) had its key exposed with 375 power traces. Results with Boolean masking are shown in Fig. \ref{fig:scatable} (b), where the key is exposed with 6200 power traces. If Boolean masking is combined with the dynamic logic implementation, Fig. \ref{fig:scatable} (c), it takes 375 k power traces, 1000x compared to the baseline, to expose the key.

Next, we investigated effectiveness of the clock randomization on the baseline microprocessor (NCM-uP). We tested three types of clock edge randomization. In Fig. \ref{fig:scatable} (d) 25\% of clock edges were skipped, and it took 850 k power traces to expose the key. The required number of traces to expose the key increased dramatically to 6.2 M, and 9.1 M, as we skipped 50\% and 75\% of the clock edges, as respectively shown in Fig. \ref{fig:scatable} (e) and (f).

Pre-processing techniques such as dynamic time warping (DTW) may be used to align power traces before a CPA attack~\cite{attackDtw2007}. The DTW algorithm originated from speech recognition systems to match spoken words to a database containing prerecorded words with different timing. We use a FastDTW variant which has complexity $O(Tk)$, where $T$ is number of traces, and $k$ is trace length~\cite{attackDtwImproved2011}. In our experiments with DTW pre-processing, the radius parameter was set to 90. Larger radius enhance representation accuracy of the original DTW algorithm, but significantly increase computing time and memory usage.

Fig. \ref{fig:scatable} (g), (h), and (i) show the CPA attack results with DTW pre-processing on the baseline microprocessor (NCM-uP), with clock randomization enabled. In Fig. \ref{fig:scatable} (g), with 25\% of clock cycles skipped, the key was exposed with only 60 k traces, which is 14.2x fewer traces compared to CPA without DTW pre-processing, as shown in Fig. \ref{fig:scatable} (d). However, it takes nearly 10x longer computation time. Similarly, Fig. \ref{fig:scatable} (h), and (i), with 50\% and 75\% of clock cycles skipped, show that the key was not exposed after a 500 k traces CPA attack using DTW pre-processing, as opposed to the 6.5 M and 9.1 M traces required to expose the key without DTW.

We also assess the information leakage of all counter-measures combined. For that, we skip 50\% of clock cycles at the BM-DDL-uP microprocessor, which implements both Boolean masking and DDL. Fig. \ref{fig:scaall} shows that CPA attacks using (a) 20 M traces without pre-processing, and (b) 2 M traces with DTW, were not enough to expose the key.

\subsection{Comparison with Prior Work}

\begin{table}[t]
    \centering
    \caption{Comparison with other secure microprocessors.}
    \label{tab:mcucmp}
    \begin{threeparttable}
        \begin{tabular}{@{}lcccc@{}}
\toprule
                         & \textbf{This work} & \textbf{CHES'07}       & \textbf{CARDIS'16}       & \textbf{DAC'19}                         \\
                         &                         & \cite{lsMDPL80512007}  & \cite{mskDomainUp2016}   & \cite{mskRambus2019}                    \\ \midrule
Technology               & 65 nm                   & 130 nm                 & FPGA                     & FPGA                                    \\
Architecture             & RISC-V                  & 8051                   & RISC-V                   & RISC-V                                  \\
Datapath                 & 1 bit                   & 8 bit                  & 32 bit                   & 32 bit                                  \\
Clock rand               & Yes                     & No                     & No                       & No                                      \\
Entropy source           & Jitter                  & No                     & No                       & No                                      \\
PRNG                     & Yes                     & Not avail              & Not avail                & Not avail                               \\
Rand bits/cycle          & 243                     & 1                      & --                       & --                                      \\
Mem addr Scr             & Yes                     & No                     & No                       & No                                      \\
Mem data enc             & Yes                     & No                     & No                       & Yes                                     \\
Masking                  & \cite{mskBiru2017}      & MDPL~\cite{lsMDPL2005} & DOM~\cite{mskDomain2016} & TI~\cite{mskThreshold2006}*             \\
Fully masked             & Yes                     & Yes*                   & No                       & Yes*                                    \\
Pre-charge logic         & DDL                     & MDPL~\cite{lsMDPL2005} & No                       & No                                      \\
Methodology              & Auto.                   & Not avail              & Manual                   & Not avail                               \\
Traces to discl.         & 375                     & 5 k                    & --                       & 15 k                                    \\
Max Traces               & 20 M                    & 300 k                  & 100 M                    & 3 M                                     \\
Attack types             & CPA/DTW                 & DPA                    & TVLA                     & TVLA                                    \\
Open-source              & Yes~\cite{resCodeSCA}   & No                     & No                       & No                                      \\ \bottomrule
\end{tabular}
        \begin{tablenotes}[para,flushleft]
            Notes: (*) few details were provided in the publication.
        \end{tablenotes}
    \end{threeparttable}
\end{table}

There are several implementations of Boolean masked microprocessors in the literature~\cite{lsMDPL80512007, mskDomainUp2016, mskRambus2019}. Table~{\ref{tab:mcucmp}} provides a comparison of our work with previous publications with respect to several key security features listed in the first column. It is clear from Table~{\ref{tab:mcucmp}} that our work covers a broad range of techniques and components necessary to implement a secure microprocessor.

Unlike other works, we used a CPU implementation with datapath size of 1~bit. Such decision was made due to area constraints in our testchip, but our methodology can be applied to any digital design, of any size. In fact, our RTL and implementation scripts are publicly available to interested researchers~\cite{resCodeSCA}. However, comparing the effectiveness of countermeasures from different publications is not trivial. Changes in the acquisition system, and the presence of combined countermeasures make it hard to draw any definitive conclusions. It is also important to mention that the effectiveness of our countermeasures will likely increase when applied to larger designs. In addition to higher switching noise, the sensitive signals of larger designs will be relatively weaker, compared to the total power consumption, requiring attackers to collect more power traces.

\section{Conclusion\label{sec:conc}}

We demonstrated a bit-serial RISC-V microprocessor implementation with no plain-text data. Our design uses Boolean masking at the logic level, and dynamic domino logic at the transistor level. We selected a a set of counter-measures that require no changes to the input RTL code. Unlike previous literature, our methodology is fully integrated with CAD tools, and can be applied to digital designs of any size or complexity. We also provided details on other key components of secure ICs, such as clock randomizer, memory protection, and random number generator. The random numbers generated with our RNG pass on all NIST tests. Side-channel analysis on the baseline implementation extracted the AES key using only 375 traces, while our secure microprocessor was able to withstand attacks using 20~M traces.

\bibliographystyle{plain}
\bibliography{sca}

\vspace{-0.4in}

\begin{IEEEbiography}[{\includegraphics[width=1in,height=1.25in,clip,keepaspectratio]{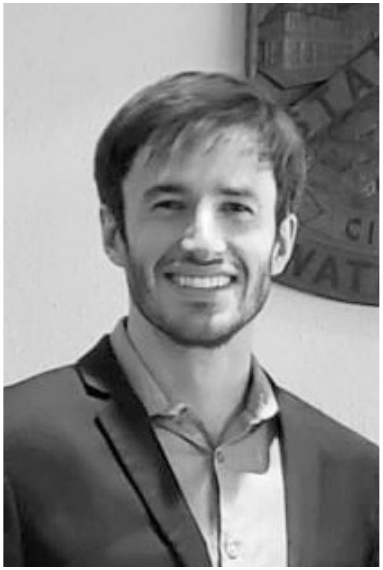}}]{Kleber Stangherlin}
Kleber received his B.Sc. in Electrical Engineering at PUCRS, and M.Sc. in Microelectronics at UFRGS, both in Brazil. He has more than 6 years of industry experience designing security focused integrated circuits. He had key contributions to the cryptographic cores and countermeasures used in the first EAL 4+ certified chip designed in the southern hemisphere. Currently, Kleber is pursuing a PhD at University of Waterloo in Canada, where he conducts research in hardware security.
\end{IEEEbiography}

\vspace{-0.4in}

\begin{IEEEbiography}[{\includegraphics[width=1in,height=1.25in,clip,keepaspectratio]{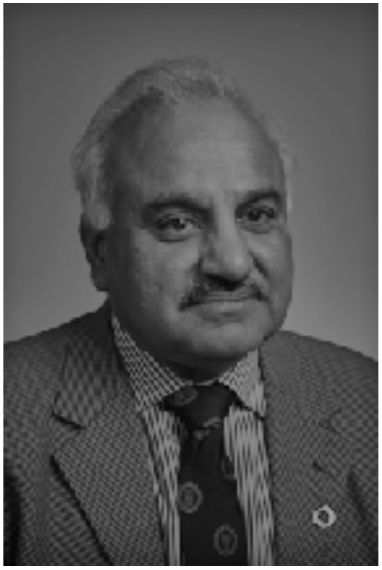}}]{Manoj Sachdev}
Manoj Sachdev is a Professor and Interim Department Chair in the Department of Electrical and Computer Engineering at the University of Waterloo. He has contributed to over 180 conference and journal publications, and has written 5 books. He also holds more than 30 granted US patents. Along with his students and colleagues, he has received several international research awards. He is a Fellow of the Institute of Electrical and Electronics Engineers (IEEE), and Fellow of the Engineering Institute of Canada. Professor Sachdev serves on the editorial board of the Journal of Electronic Testing: Theory and Applications. He is also a member of program committee of IEEE Design and Test in Europe conference.
\end{IEEEbiography}

\vfill

\end{document}